\documentclass[12pt,a4paper]{article}
\usepackage{setspace}

\usepackage{pdflscape}
\usepackage{chngcntr}
\usepackage[english]{babel}
\usepackage{multicol}
\usepackage{authblk}
\usepackage{graphicx}
\usepackage{cite}
\usepackage{amsmath}
\usepackage{booktabs}
\usepackage{setspace}
\usepackage[section]{placeins}
\usepackage{longtable}
\usepackage{rotating}
\usepackage{colortbl}
\usepackage{wrapfig}
\usepackage{makecell}
\usepackage{multirow}
\usepackage{pdflscape}
\usepackage{threeparttable}
\usepackage{threeparttablex}
\usepackage[numbers]{natbib}
\usepackage{longtable}
\usepackage{hhline}
\usepackage[resetlabels]{multibib}
\newcites{appendix}{Supplemental References}
\renewcommand\footnotemark{}
\usepackage{lineno}
\usepackage{wrapfig}
\usepackage{lscape}
\usepackage{rotating}
\usepackage{epstopdf}

\makeatletter
\def\maxwidth{\ifdim\Gin@nat@width>\linewidth\linewidth\else\Gin@nat@width\fi}
\def\maxheight{\ifdim\Gin@nat@height>\textheight\textheight\else\Gin@nat@height\fi}
\makeatother

\usepackage{blindtext}
\usepackage{subfigure}
\usepackage{lipsum}
\usepackage{booktabs}
\usepackage[margin=15pt,font=small,labelfont={bf}, singlelinecheck=false]{caption}
\usepackage{subcaption}
\usepackage[dvipsnames, table]{xcolor}
\usepackage{graphicx}
\usepackage{authblk}

\usepackage{makecell}

\usepackage[colorlinks=true,
linkcolor=Red,
urlcolor = Blue,
citecolor=Blue]{hyperref}%
\usepackage[top=1in,right=1in,left=1in,bottom=1in]{geometry}

\newcommand{\supplementarysection}{%
  \setcounter{figure}{0}
  \let\oldthefigure\thefigure
  \renewcommand{\thefigure}{S\oldthefigure}
  \section{Supplementary section}
  \let\oldchapter\chapter
  \renewcommand{\chapter}{
    \let\thefigure\oldthefigure
    \let\chapter\oldchapter
    \oldchapter
  }
}

\title{Offshoring Emissions through Used Vehicle Exports
\thanks{
\textbf{Correspondence:} All correspondence and requests for material should be directed to Dr Saul Newman, the corresponding author, at \href{mailto:saul.newman@demography.ox.ac.uk}{saul.newman@demography.ox.ac.uk}.
\textbf{Acknowledgements:} 
We would like to thank Professor Sarah Sharples for her support as chief scientific officer at the Department for Transport, climate change engineer Dr Rabee Jibrin for his wonderful feedback, Mr Grant Thunder for his excellent help and advice, and the Driver and Vehicle Standards Agency and the Department for Transport for their substantial help and support, without which this project would not be possible.
\textbf{Project Ethics:} This project was approved by the University of Oxford’s Departmental Research Ethics Committee (Sociology) under ethics approval SOC\_R2\_001\_C1A\_21\_66.  A compliance assessment was undertaken by Oxford Population Health. \textbf{Replication Materials:} Imputation models and code are freely available under a creative commons CC-BY-NC 4.0 licence from a \href{https://doi.org/10.6084/m9.figshare.23702571}{figshare repository} or on request from the corresponding author. \textbf{Funding:} The authors are grateful for support from the Leverhulme Trust (Grant RC-2018-003) for the Leverhulme Centre for Demographic Science, and Nuffield College.
}
}

\author[1, 2]{Saul Justin Newman}
\author[3]{Kayla Schulte} 
\author[1, 4, 5]{Micol Matilde Morellini}
\author[1, 4]{Charles Rahal} 
\author[1, 4]{Douglas R. Leasure}

\affil[1]{ Leverhulme Centre for Demographic Science, University of Oxford}
\affil[2]{ University College, University of Oxford}
\affil[3]{ Imperial College London}
\affil[4]{ Nuffield College, University of Oxford}
\affil[5]{ Department of Sociology, University of Oxford}

\begin{document}

\maketitle 
\vspace{-0.25in}
\begin{center}
\normalsize{\textbf{Please cite this article as:}\\ \vspace{.05in}
Newman, S.J., Schulte, K., Morellini, M.M., Rahal, C., and Leasure, D., `Offshoring emissions through used vehicle exports'. \textit{Nat. Clim. Chang.} (2024). https://doi.org/10.1038/s41558-024-01943-1}\\ \vspace{.2in}
\end{center}
\begin{abstract}
Policies to reduce transport emissions often overlook the international flow of used vehicles. We quantify the rate at which used vehicles generated CO$_\textrm{2}$ and pollution for all used vehicles exported from Great Britain – a globally leading used-vehicle exporter – across 2005-2021. Destined for low-middle income countries, exported vehicles fail roadworthiness standards and, even under extremely optimistic ‘functioning-as-new’ assumptions, generate at least 13-53\% more emissions than scrapped or on-road vehicles.
\end{abstract} \vspace{.05in}


\newpage
\section*{Main Article}
Transport is the largest emitting sector of greenhouse gases, accounting for a quarter to a third of all emissions in developed countries \citep{2a701673-en, wallington2014n2o} with serious consequences for both climate and health \cite{anenberg2017impacts, lelieveld2015contribution, cohen2017estimates}. Air pollutants such as nitrogen oxides (NOx), which are effectively reduced when standards are enforced, cause millions of deaths each year \citep{cohen2017estimates, babiker2005climate}. These impacts fall unequally on lower-middle income countries (LMICs, \citep{babiker2005climate}) which suffer more overall and per-capita pulmonary deaths from air pollution \citep{lelieveld2015contribution, cohen2017estimates, fuller2022pollution} and stand to suffer the greatest impacts from climate change \citep{mertz2009adaptation}. 
The source of vehicles in LMICs is dominated by unregulated trade \citep{2a701673-en, wallington2014n2o, anenberg2017impacts, baskin20used, davis2010international}. As of 2020, 100 countries receiving used vehicles had no vehicle emissions standards \citep{baskin20used}, and only eleven had `very good' \citep{baskin20used} regulations. However, the USA, EU, Japan, and UK collectively supply 90\% of used vehicles exported to (non-EU) LMICs \citep{baskin20used}. The potential for rapid regulation is therefore incumbent on just four jurisdictions, all of which already maintain high vehicle emission standards. Using comprehensive government databases, we quantify per-kilometre rates at which vehicles generate carbon and pollution for every vehicle (N=6,921,292) legally exported from Great Britain between January 2005 and December 2021. We compare these vehicle emissions to every private vehicle driven in Great Britain during the same period, and those that would have been driven if they had not been scrapped. 

\begin{figure}[!t]
\begin{center}
    \includegraphics[width=1\textwidth]{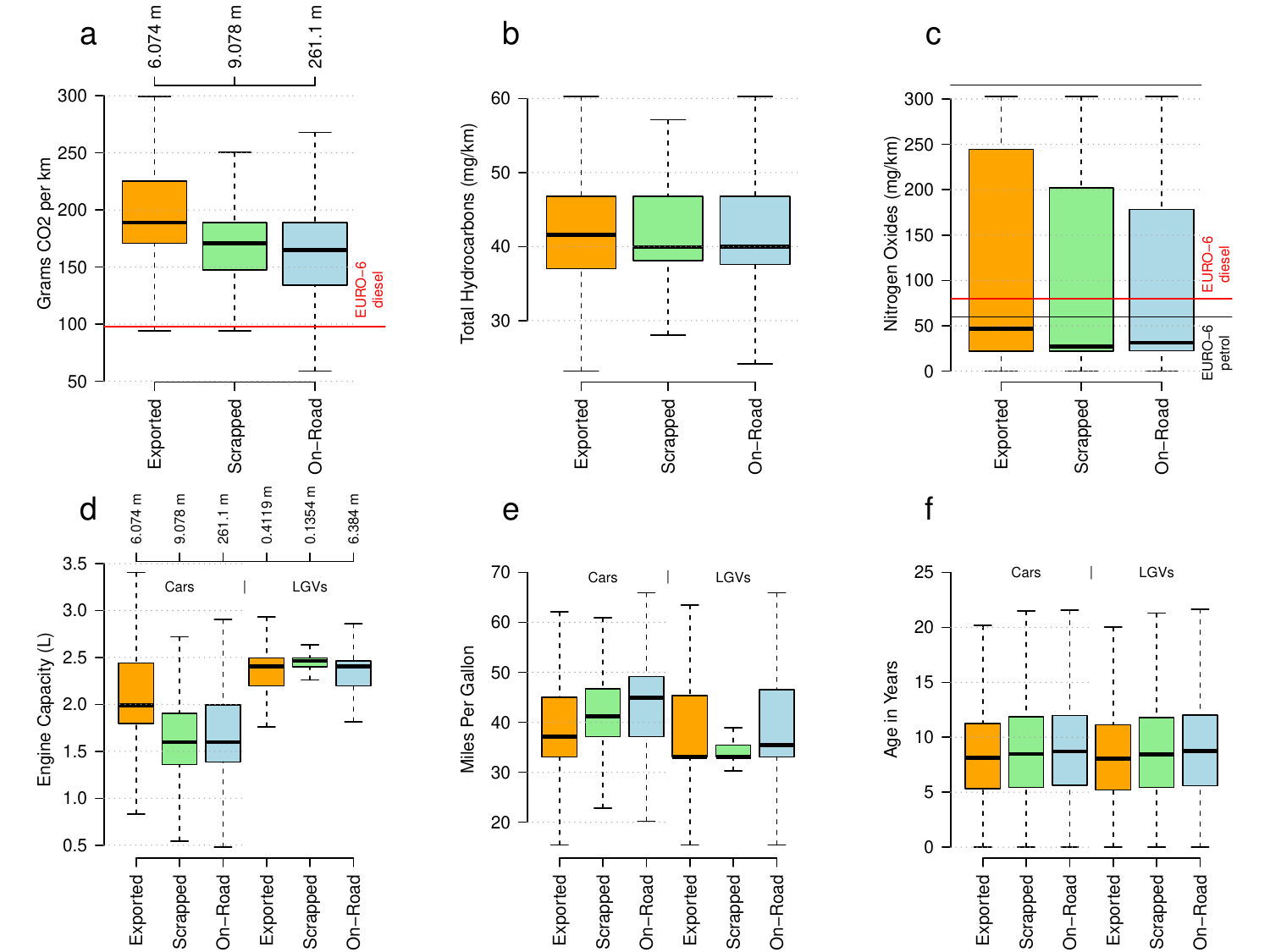}
    \caption{Pollution generated per kilometre by used vehicle fleets. Exported cars (orange) generate at least 13-24\% more CO$_\textrm{2}$ per kilometre (a), similar average but (through overdispersion) higher median fine particulate matter pollution (b), and (c) at least 53\% more nitrogen oxides per kilometre than scrapped (green) or on-road (blue) vehicle fleets. Both exported car and LGVs fleets also have larger observed engine capacities (d) and worse fuel efficiency (e) despite being similar ages to on-road fleets and much younger than contemporary scrapped vehicles (f). Boxes show interquartile range (IQR), whiskers are 1.5x IQR, sample size (millions) on supplementary x-axes.} \label{fig_1}
    \end{center}
\end{figure}

These data reveal substantially higher rates of CO$_\textrm{2}$ and pollution generation in exported vehicles, even under optimistic `functioning-as-new’ emissions intensity estimates that assume no vehicle modifications or vehicle degradation with age (Fig \ref{fig_1}). Exported cars generate at least 23g (13\%) more CO$_\textrm{2}$ per kilometre than cars scrapped in the same period (Fig \ref{fig_1}a; \cite{newman_figshare}), and at least 29g (17\%) more CO$_\textrm{2}$ per kilometre than the contemporary on-road used car fleet (Fig \ref{fig_1}a; mean 197.0, 174.4, and 168.6g/km CO$_\textrm{2}$ for exported, scrapped, and on-road fleets respectively; interquartile ranges IQR 170.1-225.3, 147.5-189.1, and 134.2-188.9g/km). Emissions figures were even more striking for other pollutants. Exported cars emit similar amounts of hydrocarbon particulates (Fig \ref{fig_1}b) but 48mg/km more NOx (53\% higher; Fig \ref{fig_1}c) than scrapped cars. Likewise, observed engine capacities were larger (Fig \ref{fig_1}d) and fuel efficiency at least 9\% worse, by 3.3 miles per gallon (MPG, mean 38.5, 41.8 and 44.4mpg for exported, scrapped and on-road used fleets respectively; IQR 33-45mpg exported, 37-47mpg scrapped, and 37-49mpg on-road fleets). A substantial fraction (42\%) of exported diesel vehicles were predicted to fail the current EURO-4 emissions standards \citep{eurlex} that form the legal roadworthy minimum for all vehicles registered after 2000. A surprising 83\% were predicted to fail the EURO-6 diesel \citep{eurlex} CO$_\textrm{2}$ emissions standards, and 98\% failed the EURO-6 carbon monoxide and NOx standards. These differences are not the result of over-dispersion, where a few high-emitting exports \citep{eurlex} dragged up the average: similar or even larger gaps in pollution rates were observed for the median pollution rates of exported, scrapped, and contemporary on-road used vehicles (Fig \ref{fig_1}).

\begin{figure}[!t]
  \centering
    \includegraphics[width=\textwidth]{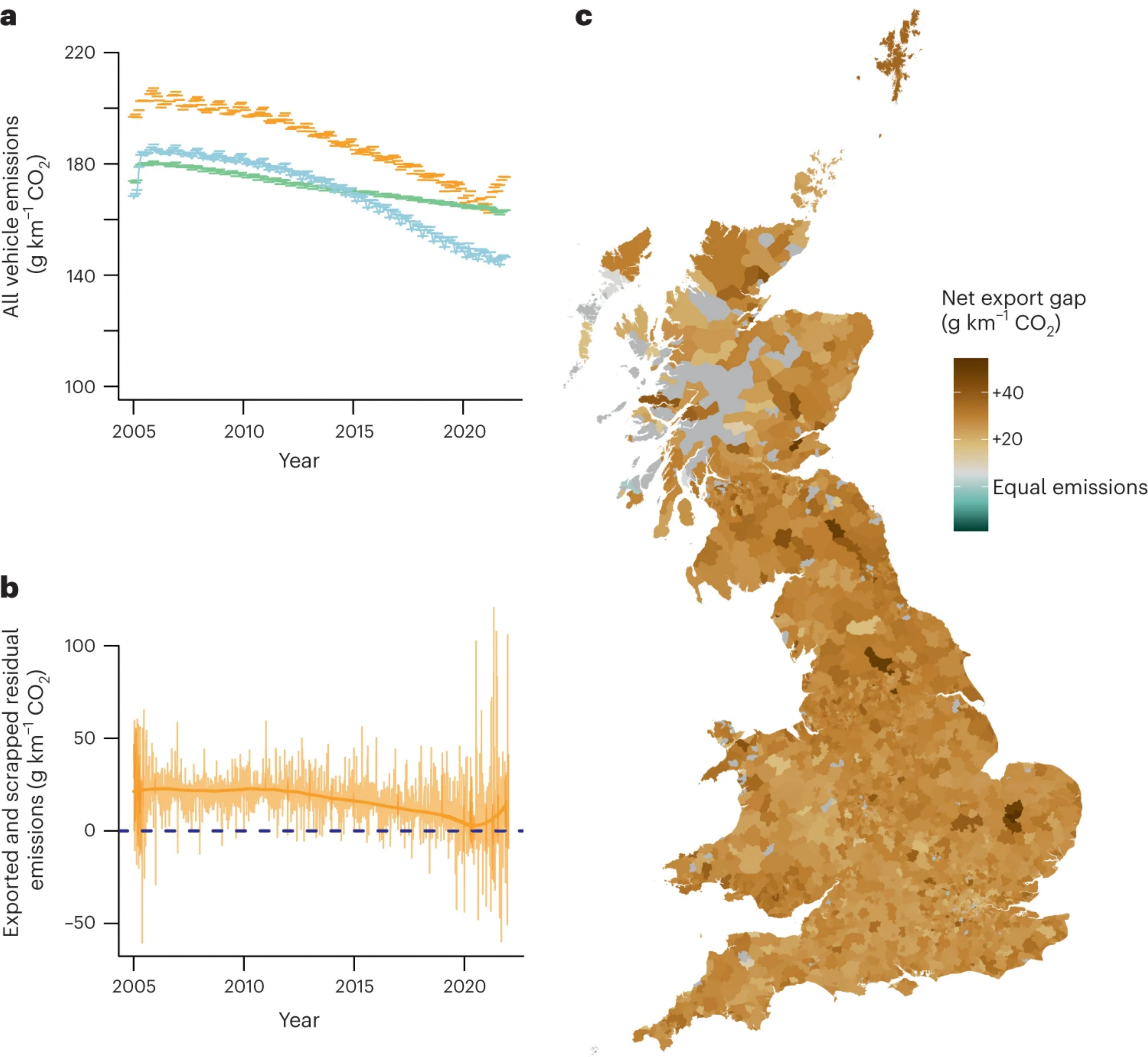}
  \caption{Export of consistently higher-emitting vehicles from Great Britain. Exported vehicles (orange) are consistently more polluting (a) than contemporary scrapped (green) and used on-road (blue) vehicles. Cleaner choices and improving standards have lowered exported emissions, after considerable delay, but the gap between exported and scrapped fleets persists (b) and is growing post-pandemic. Scrapped vehicles are cleaner than exports across 5,874 (95\%) of our 6,145 observed postcode regions (2005-2021 inclusive) covering every community in Britain (c). Grey regions have insufficient data or no inspection sites. Orange line (b) shows locally weighted smoothed spline; note non-zero y-axis (a) to emphasise variation.
}\label{fig_2}
\end{figure}

Daily-resolution data for six million exported cars reveals that the gap between exported, on-road, and scrapped fleets is consistent over time (Fig \ref{fig_2}a), apart from a narrowing and then rapid expansion of this gap over 2020-2021 alongside distortions of trade patterns, used car prices, and vehicle testing regulations during the COVID-19 pandemic (Fig \ref{fig_2}b). That is, Great Britain persistently scraps lower-emissions vehicles while exporting higher-emission vehicles (Fig \ref{fig_2}b). While geographical disparities due to uneven concentrations of upmarket export vehicles or differing ‘on-road’ usage were anticipated, this export gap was remarkably uniform. Almost every British postcode region (95\%), representing a full cross-section of society, export higher-polluting vehicles than those they drive or scrap (Fig \ref{fig_2}c). Exported vehicles will likely generate more pollution per kilometre independent of their destinations and patterns of use, for simple physical reasons: compared to their more-efficient scrapped alternatives, export vehicles have larger observed engine capacities (Fig \ref{fig_1}d) and lower operating efficiencies (Fig \ref{fig_1}e), despite a younger average age (Fig \ref{fig_1}f). These fixed factors also mean that degradation rates are likely rank-conserved, especially for CO$_\textrm{2}$, and high-polluting vehicles will remain the most polluting as vehicle fleets age. 

Air quality outcomes are more nuanced than individual emission categories, including how vehicles are driven, road conditions, engine age, climate, payload, and maintenance schedules \cite{bernard2020development, greene2017does}. Vehicles also generate pollution, such as ozone or non-tailpipe emissions, for which testing data was not available. Observational data are urgently needed to fill this gap. Adding to the challenges of measuring emissions, emissions testing data have long been manipulated. For example; `Dieselgate' involved nine major manufacturers using `defeat devices' to alter performance and intentionally deceive environmental agencies and regulators. In the Dieselgate aftermath, vehicle manufacturers are, incredibly, allowed to legally manipulate vehicles during new-car emissions testing \citep{greene2017does, muncrief2016defeat} by, for example, removing wing mirrors and seats, taping up high-drag surfaces, or hard-baking and over-inflating tyres. Manufacturers are also now allowed to `adjust’ emissions estimates \citep{muncrief2016defeat} by 4.5\% and program vehicles to turn off emission-reduction devices \cite{DEFEATER} when the weather becomes `too hot’ or `cold' \citep{DEFEATER}. Manufacturers define `hot' and `cold'. For example, Renault told the French government their emissions control devices should shut off above 35°C and below 17°C (Paris is colder 83\% of the time \citep{muncrief2016defeat}) to `protect the engine’ \citep{muncrief2016defeat}, at the cost of protecting the climate and human health. As a result, real-world emissions increasingly overshoot (currently by ~50\%) emissions measured during testing \citep{greene2017does, ICCT2015, europe2017road}. Some 13\% of diesel cars in the EU now emit NOx at over ten times the legal standard \cite{europe2017road}, outnumbering the 10\% that actually meet those standards. 

The lack of emissions standards in most destination countries also results in the routine stripping of emission-reduction devices for resale \cite{HETI} or melting down before export \citep{baskin20used, HETI}. One study tested 160 vehicles destined for Africa from the EU \cite{HETI}. Of the vehicles that could start, 85-93\% failed to meet the (roadworthy-minimum) EURO-4 emissions standards, 20\% of petrol (gasoline) vehicles did not comply with any emission standard at all \citep{HETI}, and 10\% had their catalytic converters cut out \cite{HETI} increasing NOx and carbon monoxide pollution 10-fold. Used vehicles were exported at an average 8.5 years of age (IQR 5.0-10.6 years) and both emissions \citep{bernard2020development} and fuel efficiency \citep{greene2017does} degrade with age. Our models, therefore, likely underestimate vehicle pollution rates substantially, by relying on new car testing data that does not account for increasing emissions generation from vehicle aging or modifications. The nonstationarity and complexity of emissions degradation curves \citep{bernard2020development, greene2017does} and vehicle modifications mean that direct measurements -- like those increasingly captured by annual vehicle emissions tests required in the UK -- are needed to improve estimates of exported emissions. Actual emissions are likely far higher \citep{bernard2020development, greene2017does, europe2017road}, perhaps 150\% higher for CO$_\textrm{2}$ under ideal `European-style' driving conditions \citep{europe2017road}, but enormous and unnecessary gaps in our knowledge remain.

As with carbon leakage from heavy industry and manufacturing \citep{babiker2005climate}, rich countries appear to be offshoring the cost of replacing high-polluting vehicles. There are, however, some positive trends. While many improved emissions are artefacts of manipulated testing \citep{DEFEATER, ICCT2015, reynaert2016corrective}, better fuel efficiency and air quality standards in the USA, UK, EU, and Japan are slowly reducing the estimated pollution of exported vehicles over time (see Fig \ref{fig_2}a-b for Great Britain). These four jurisdictions are the collective source of over 95\% of light used vehicle exports worldwide and, despite creating 40\% of global transport emissions \citep{crippa2022co2}, implement world-leading vehicle emissions standards inside their own borders. Imparting the same standards on exported vehicles, preventing the removal of emission-reduction devices, and redirecting clean vehicles from the scrapyard to the export fleet would all positively impact global emissions. Export licences can be indexed to increase duties on dirty vehicles or subsidise clean vehicle exports. Export countries have very few major vehicle ports \citep{2a701673-en} and thousands of mechanics qualified to evaluate legal roadworthy standards. Tasking mechanics to randomly spot-check vehicles in port, and issue penalties when vehicles fail emissions tests, would be an extremely low-cost intervention to stem the dirty used vehicle trade. Such measures would not necessitate increased vehicle prices, which can reduce access to the economic benefits of vehicle ownership. Supply shocks can be mitigated or avoided by using policy and incentives, and by redirecting clean vehicles from scrapyards to export. Cleaner vehicles also have smaller and more fuel-efficient engines on average, and lower ongoing costs \citep{miotti2016personal} over the life of the vehicle, reducing net economic burdens. Potential short-term price increases imposed on individuals are also offset by long-term reduction in the societal and economic costs from pollution \citep{wallington2014n2o, lelieveld2015contribution, cohen2017estimates} and climate change \citep{2a701673-en, anenberg2017impacts, mertz2009adaptation}. Most LMICs are placing this consideration above others, with widespread moves to ban the import of dirty used vehicles, regardless of price shocks \citep{baskin20used, miotti2016personal}. These policies reflect a growing desire for clean air over cheap cars. However, such moves are struggling for traction due to a lack of policing and resources, and the unstemmed flow of unregulated imports \citep{miotti2016personal}.
 
Developed economies can aid these goals and reduce the damage from vehicular emissions by raising export standards to match their own internal legal minimum standards. Such low-cost interventions are an immense opportunity for rich high-emitting countries to reduce global emissions and cut pollution in the developing world. To instead overlook this problem, and allow the continued flow of high-emissions vehicles, would be a devastating missed opportunity and an ethical failure.

\section*{Online Methods}

Data were obtained from the Department for Transportation – a department of the government of the United Kingdom – for all 65 million privately registered used vehicles undergoing mandatory annual vehicle inspections. Traditionally termed `MOT’ tests, they were undertaken across in the UK between 1 January 2005 and 31 December 2021. Used vehicles were defined as all vehicles that had undergone at least one prior inspection. These annual vehicle inspections are required by law in the UK to assess road worthiness. They begin one year after the vehicle is first registered for motorcycles and scooters, and three years after first registration for all other vehicles (in Northern Ireland, the equivalent requirement applies after four years). Alongside these data, we obtained linked vehicle-specific data on all used vehicles that had been scrapped or issued a certificate of destruction (N=9,077,804), and all vehicles that had been flagged as exported (N = 6,922,292). Both export and scrappage certifications are supplied with exact dates.

Used-vehicle summary statistics for predicted emissions, age, observed engine capacity, and all other vehicle properties were calculated at a daily resolution, from 1 January 2005 - 31 December 2021 inclusive, for a comprehensive sample of every on-road vehicle test (267.5 million tests; one random test per annum was selected for each vehicle), and for every scrapped (issued a vehicle scrappage certificate or certificate of destruction) or exported vehicle \cite{newman_figshare}. We restricted our `on-road' data to one randomly sampled roadworthy test per vehicle per year to avoid the oversampling of mechanically unreliable vehicles, which are re-tested each time they fail a roadworthy certificate \cite{newman_figshare}. This resulted in predictions of emissions intensity across 261.1 million vehicle-days, or forty-two thousand vehicles per day, for cars (or class 4 vehicles) known to be on-road at the time of their roadworthy inspection. 

All analyses were performed using R version 4.0.5 (see Supplementary Code). Some 993 vehicles (0.015\%) that were erroneously flagged as both exported and scrapped or destroyed were removed from analysis. As the date of first use, and the date of each vehicle inspection, scrappage, destruction, or export were reported, we calculated vehicle ages at each of these events. Some 60,642 (0.4\%) of the reported dates were `impossible' and were excluded, as they were in potentially reverse order, or contained typographic errors. Another 7,380 vehicles over the age of 110 years (0.05\%) were excluded from analysis as they largely -- but not entirely -- constituted age-related coding errors.

The vehicle inspection and scrappage data were matched to a public dataset of fuel efficiency and emissions data from the Vehicle Certification Agency, for seventy thousand measurements tabulated by vehicle make, model, fuel type, and year, measured across vehicles sold in the UK from 2000 onwards. Emissions data for carbon dioxide, carbon monoxide, nitrogen oxides, total hydrocarbon particulates and fuel efficiency of UK vehicles were also obtained from the Vehicle Certification Agency. Given the near-complete lack of emissions testing data for motorcycles and other vehicle classes, emissions testing data were only captured for cars and vans, and were curated, quality-controlled, and matched to the Department for Transport data \cite{newman_figshare}. 

Matching these emissions testing data resulted in 1.3 million exactly matched CO$_\textrm{2}$ measurements for cars in the exported or scrapped fleet (8.4\% of all exported cars and 8.1\% of all scrapped cars) but only 3222 exact-matched CO$_\textrm{2}$ measurements for LGVs (0.3\% of all LGVs), a number largely restricted by the abundance of rarer or untested models, and the difficulty in exact-matching heterogeneous make and model descriptions provided by the Vehicle Certification Agency. Exported vehicles with exact matches for measured emissions were lower (just 678 or 0.16\% of all exported LGVs), excluding the reliable imputation of emissions from LGVs. 

These data highlighted the extensive need for better, more comprehensive emissions testing for both new, on-road, and exported vehicles. The regulatory environment needs to be restructured to fill these gaps. However, while testing regimes and data were insufficient to impute on-road emissions in motorcycle or LGV fleets, accurate imputation of car emissions and pollution was possible under the assumption that new-car testing data would remain rank-conserved over time. This assumption generates a lowest-possible estimate of emissions rates, under the assumption that used vehicles are ‘functioning-as-new’ at the point of scrappage, export, or testing. This overly optimistic assumption is a key limitation of the study, one that highlights the need for far better measurement and testing of real-world emissions across all vehicle fleets.

Imputation models were constructed to capture the emissions of all exported (N = 6,072,730), scrapped (N = 9,077,804) or on-road (N = 261.1 million tests) class 4 vehicles (cars and vans below 3T) that passed quality controls. We used the reported model year, fuel type (e.g., ‘Petrol/Gasoline’, ‘Diesel-Electric’), and engine capacity (in cubic centimetres) from the annual vehicle inspection data to construct a model for each pollution type and vehicle property. Pollution types and vehicle properties which were imputed include CO$_\textrm{2}$ (grams per km), total nitrous oxides (NOx; mg per km), total particulate hydrocarbons (mg per km), carbon monoxide (mg per km), and fuel efficiency in ‘miles per gallon’ (MPG). Imputation models were kept deliberately simple, as we predict values across the broadest possible range of vehicles using variables that were reported at almost every vehicle inspection. This avoided over-fitting and, as rare vehicles share engines and emission technology with common makes and models, achieved accurate (see Supplementary Materials: Fig \color{red}S1\color{black}) imputation for a very large ($>$99\%) fraction of cases. 

We developed imputation models using recursively partitioned regression trees \citep{therneau2015package, zeileis2008model}, a foundational and interpretable machine learning heuristic suited to discrete effects and small variable sets (Fig \color{red}S1\color{black}). This was implemented using the `rpart' package \citep{therneau2015package}. Models were trained on three input variables -- engine size in cubic centimetres, the year of vehicle manufacture, and fuel type -- using ten-fold random-sample cross-validation, with the `cp' model complexity parameter set to 0.001, and a minimum of 100 vehicle make-models used for a parent node (i.e., the `minbucket' parameter). A minimum of 25 vehicle make-models were used for a child or leaf nodes (the `minsplit' parameter; see \cite{newman_figshare}).

These models achieved high accuracy, approaching the test-retest accuracy of the Vehicle Certification Agency testing regime (Fig \color{red}S1\color{black}). For example, consecutive tests of the same vehicle for CO$_\textrm{2}$ were correlated by $r$=0.9, while our recursively partitioned regression model attained an accuracy of 
$r$=0.94 when imputing CO$_\textrm{2}$ for a random holdout sample (random 20\% holdout sample; N = 6,708 unique vehicle makes and models; Fig \color{red}S1\color{black}a-b). We found that grid-searching to further optimise model fit was therefore unnecessary, as our initial model parameters generated models that approached the highest achievable accuracy. All resulting recursively partitioned regression models, code, and imputation data are provided in the Supplementary Materials and in open, stable repositories \cite{newman_figshare}. 

\bibliographystyle{plain}
\bibliography{references}

\def\thesection{S\arabic{section}}
\setcounter{figure}{0}
\setcounter{section}{0}
\renewcommand{\figurename}{Figure}
\renewcommand{\thefigure}{S\arabic{figure}}
\newpage
\section{Files within the figshare repository}

Several supplementary files are available from the figshare repository cited in the main text:
\begin{enumerate}
    \item This document which summarises the supporting material:\\

    \item A file containing summary statistics for Figure 1, and both spatial and temporal aggregates of fleet emissions.\\

File names:\\
  	
   \quad \quad `\texttt{Figure\_data\_and\_aggregates.xlsx}'\\

\item A file summarising the accuracy of the recursively partitioned decision trees:\\
    
File names:\\

\quad \quad `\texttt{accuracy\_tables.csv}'\\

\item Four related .pdf files detailing the exact partitions of the aforementioned trees, and a zip file containing the actual models as R data serialisation files (\texttt{.RDS} files):\\

File names:\\

\quad \quad `\texttt{co2\_decision\_tree.pdf}'

\quad \quad `\texttt{co\_decision\_tree.pdf}'

\quad \quad `\texttt{mpg\_decision\_tree.pdf}'

\quad \quad `\texttt{nox\_decision\_tree.pdf}'

\quad \quad `\texttt{Models\_shareable.zip}'

\end{enumerate}

The figshare repository is available from: \\

 \quote{[19] Saul Newman, Kyla Schulte, Micol Morellini, Charles Rahal, and Douglas Leasure. Replication Materials to Accompany ‘Offshoring Emissions through Used Vehicle Exports’. \textit{figshare}, \textit{https://doi.org/10.6084/m9.figshare.23702571}, 2024.}

\newpage

\section{Supplementary Figures}

\begin{figure}[!h]
\begin{center}
    \includegraphics[width=1\textwidth]{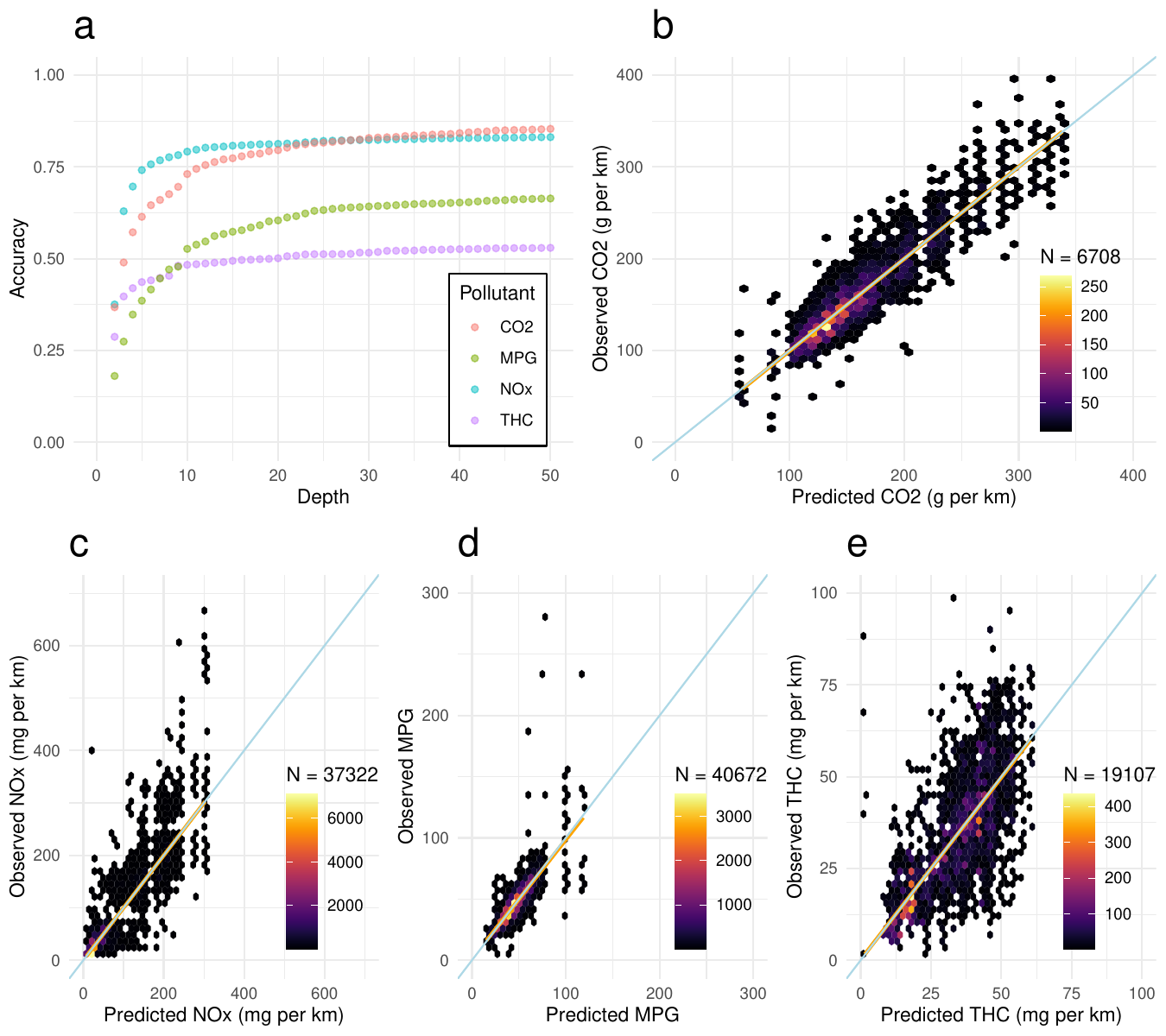}
    \caption{Imputation accuracy for randomly sampled emissions data. Emissions of non-electric vehicles were accurately imputed from recursively partitioned regression models (Supplementary Code). Predictions made on a masked random 20\% sample of vehicle emissions tests (N= 6,708 holdout make – model – year - engine combinations) showed a high degree of accuracy under ten-fold cross-validation (a; R2 on y-axis). Accuracy increased with tree depth to achieve moderate to high accuracy across CO2 (b), nitrogen oxides (NOx; c), and miles per gallon (MPG; d) prediction models, with the lowest cross-validation accuracy for total hydrocarbon emissions (THC; e). Model and manufacturer effects accounted for minimal variance, independent of these effects, and were not fit to allow imputation of rare makes and models. Filled circles in (a) are red for CO2, green for MPG, blue for NOx, and mauve for THC.} \label{fig_si_1}
    \end{center}
\end{figure}

\newpage

\section{Vehicle Testing Classes}

Vehicles fell under different standardized MOT testing classes denoting their legal status, variously, as motorcycles or scooters up to (Class 1) and over (Class 2) an engine capacity of 200cc, three-wheeled vehicles up to 450kg (Class 3), cars (Class 4; by far the most common), private vehicles and ambulances with over 13 seats (Class 5), and light goods vehicles between 3000-3500kg (LGVs; Class 7). Class 4 contained miscellaneous classes of very low-frequency vehicle types, such as three-wheeled vehicles over 450kg and ambulances, and vans or goods vehicles below 3000kg tare weight.

\newpage

\section{Acronyms}

Acronyms for Variables, Column Headers, and Code:

\begin{itemize}

\item \textbf{NOx} – Nitrogen Oxides.
\item \textbf{MPG} – Fuel efficiency in Miles per Gallon.
\item \textbf{CO2} – Carbon dioxide.
\item \textbf{THC} – Total hydrocarbons (particulate emissions).\\

\item \textbf{CP} – Complexity Parameter.
\item \textbf{nsplit} – Total number of splits in recursively partitioned tree.
\item \textbf{rel error} – Relative error of the tree.
\item \textbf{xerror} – Error rate of the tree under random-sample cross-validation.
\item \textbf{xstd} – Standard error of the tree under random-sample cross-validation.\\
\end{itemize}

\noindent Variable names for decision trees:

\begin{itemize}
\item \textbf{ModelYea} – Model calendar year of the tested vehicle
\item \textbf{EngineCa} -  Observed engine capacity, in cubic centimetres
\item \textbf{Fuel\_typ} – The Department of Vehicles Standards Agency fuel type code, classified into:
    \begin{itemize}
		\item \textbf{DIE} – Diesel
		\item \textbf{PET} – Petrol / Gasoline
		\item \textbf{CNG} – Compressed Natural Gas 
		\item \textbf{ELD} – Hybrid (Electric + Diesel)
		\item \textbf{HYB} – Hybrid (Electric + Petrol/Gasoline)
		\item \textbf{LPG} – Liquid Petroleum Gas
  \end{itemize}
\end{itemize}

\newpage

\section{Parameterising RPart Decision Trees}

All analyses were performed using R version 4.0.5 (see Supplementary Code), and in particular, we used the ‘rpart’ library (version 4.1.19). We chose the same four key parameterisations for our models (in parentheses):

\begin{itemize}

\item \textbf{cp}: prune all nodes with a complexity less than cp (cp=1e-04).

\item \textbf{minbucket}: the minimum number of observations in any terminal <leaf> node. If only one of minbucket or minsplit is specified, the code either sets minsplit to minbucket $\times$3 or $\frac{\textrm{minsplit}}{3}$, as appropriate (minbucket=100).

\item \textbf{minsplit}: the minimum number of observations that must exist in a node in order for a split to be attempted (minsplit=25).

\item \textbf{xval}: number of cross-validations (xval=10).
\end{itemize}

\newpage 

\section{How to Read Decision Trees}

Decision trees are machine-generated interpretable models and can be represented as branching flow-charts, like organisational flow charts, with a fixed structure. To read a decision tree, begin at the top node in the centre of the page. There will be a variable name and a simple decision; is the engine capacity (‘CC’) greater than 2194? Is the car a Diesel? Is the model year (‘Year’) of the car being tested greater or equal to 2010? If answering yes at any node, you always progress to the value on the left branch. On the other hand, answering ‘no’ will always take you to the right-hand node. After making each decision, the model or the user simply moves down to the next decision in the tree until they reach the bottom of the tree. From the numbers below each node, we can see how many vehicle tests these predictions are based on and the percentage of the initial vehicle tests that fall into this category. Consider the example below. If the Year is greater than or equal to 2019, we assign to that observation 24 miles per gallon. If it is less than 2019, the model then considers the engine capacity. If it’s less than 3891, the car is assigned 25 miles per gallon. If it’s more than or equal to 3891, it’s assigned 30 miles per gallon.

\begin{figure}[!h]
\begin{center}
    \includegraphics[width=0.3\textwidth]{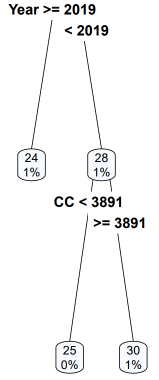}
    \label{fig_si_2}
    \end{center}
\end{figure}

These models are presented here to aid users that are not familiar with the R programming language. For anyone who is familiar with R, we provide complete, estimated models in the repository associated with this work, as we are unfortunately unable to share the associated raw data.

\newpage
\newgeometry{top=5mm, bottom=5mm, left=0mm, right=0mm}
\begin{landscape}
\begin{figure}[ht]
\begin{center}
    \includegraphics[width=1.25\textwidth]{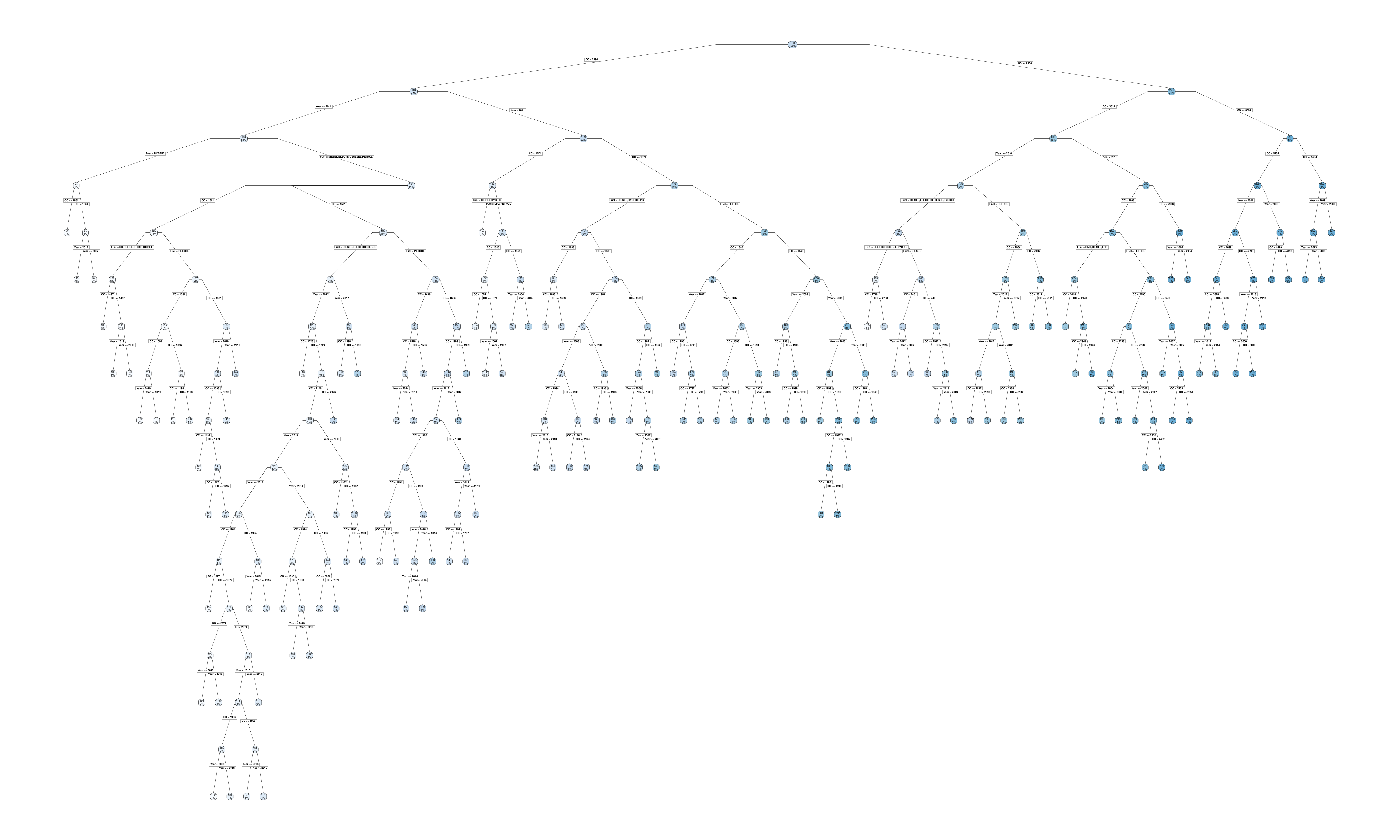}
    \caption{Recursively Partitioned Decision Tree: CO$_\textrm{2}$}
    \end{center}
\end{figure}
\end{landscape}

\newpage

\begin{landscape}
\begin{figure}[ht]
\begin{center}
    \includegraphics[width=1.25\textwidth]{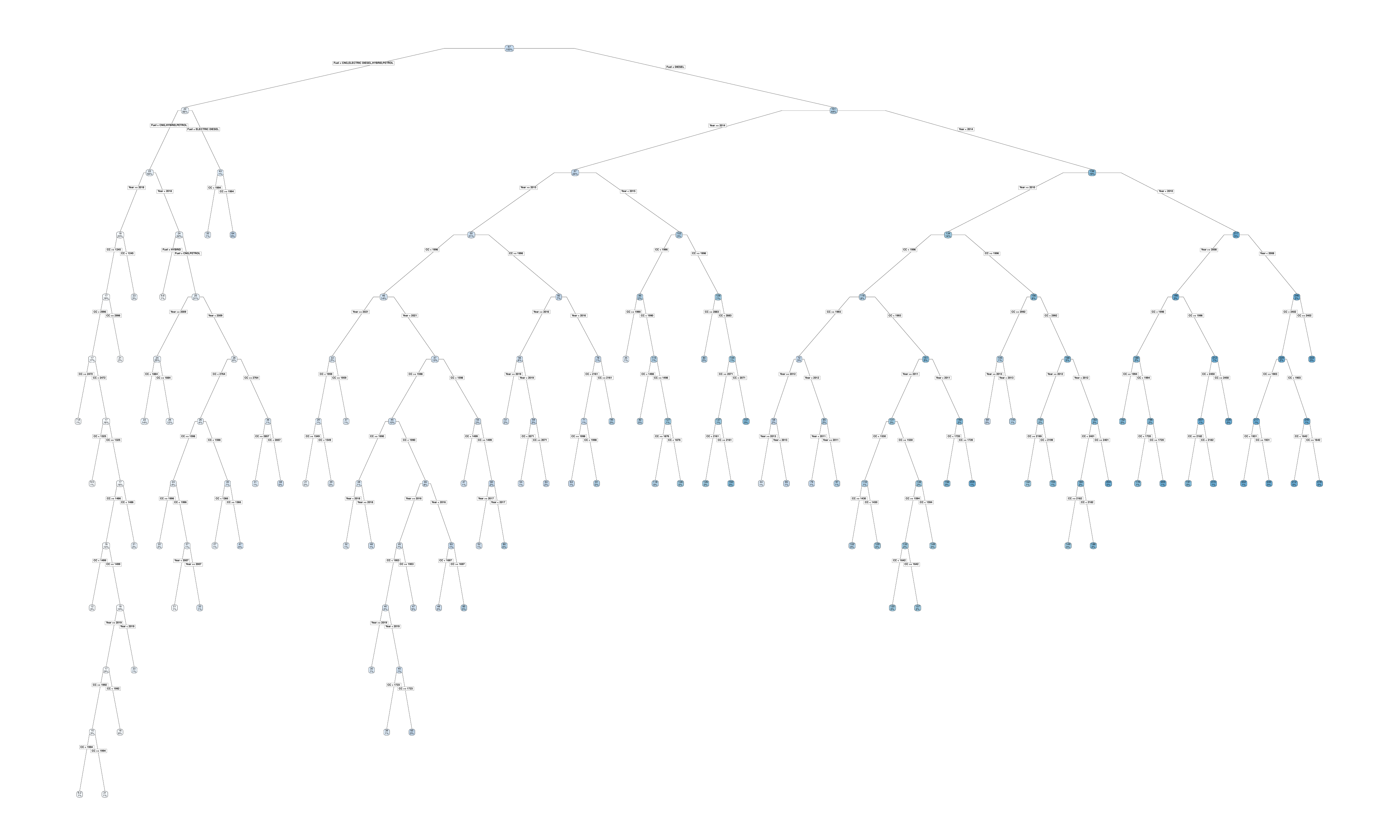}
    \caption{Recursively Partitioned Decision Tree: CO}
    \end{center}
\end{figure}
\end{landscape}

\newpage

\begin{landscape}
\begin{figure}[ht]
\begin{center}
    \includegraphics[width=1.25\textwidth]{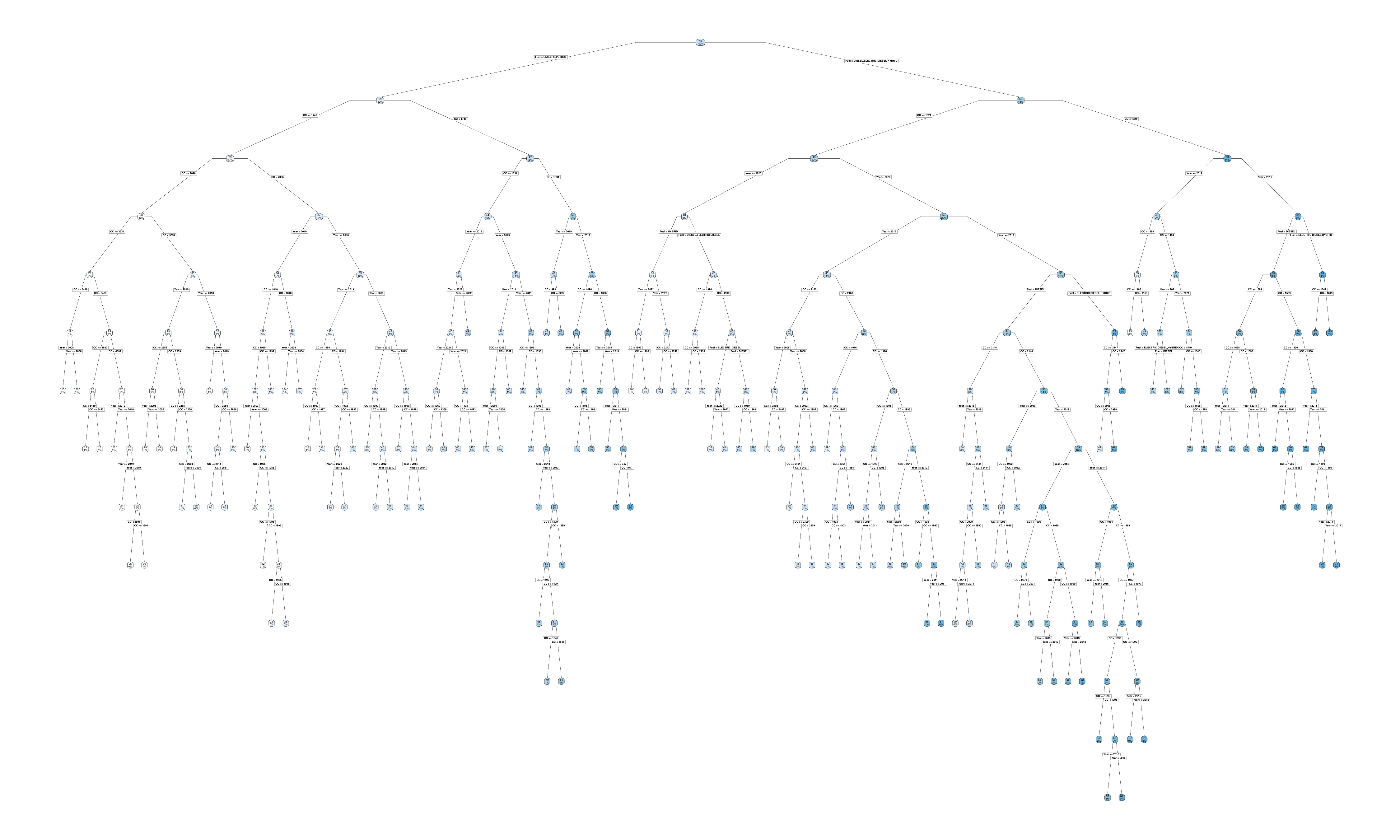}
    \caption{Recursively Partitioned Decision Tree: MPG}
    \end{center}
\end{figure}
\end{landscape}

\newpage

\begin{landscape}
\begin{figure}[ht]
\begin{center}
    \includegraphics[width=1.25\textwidth]{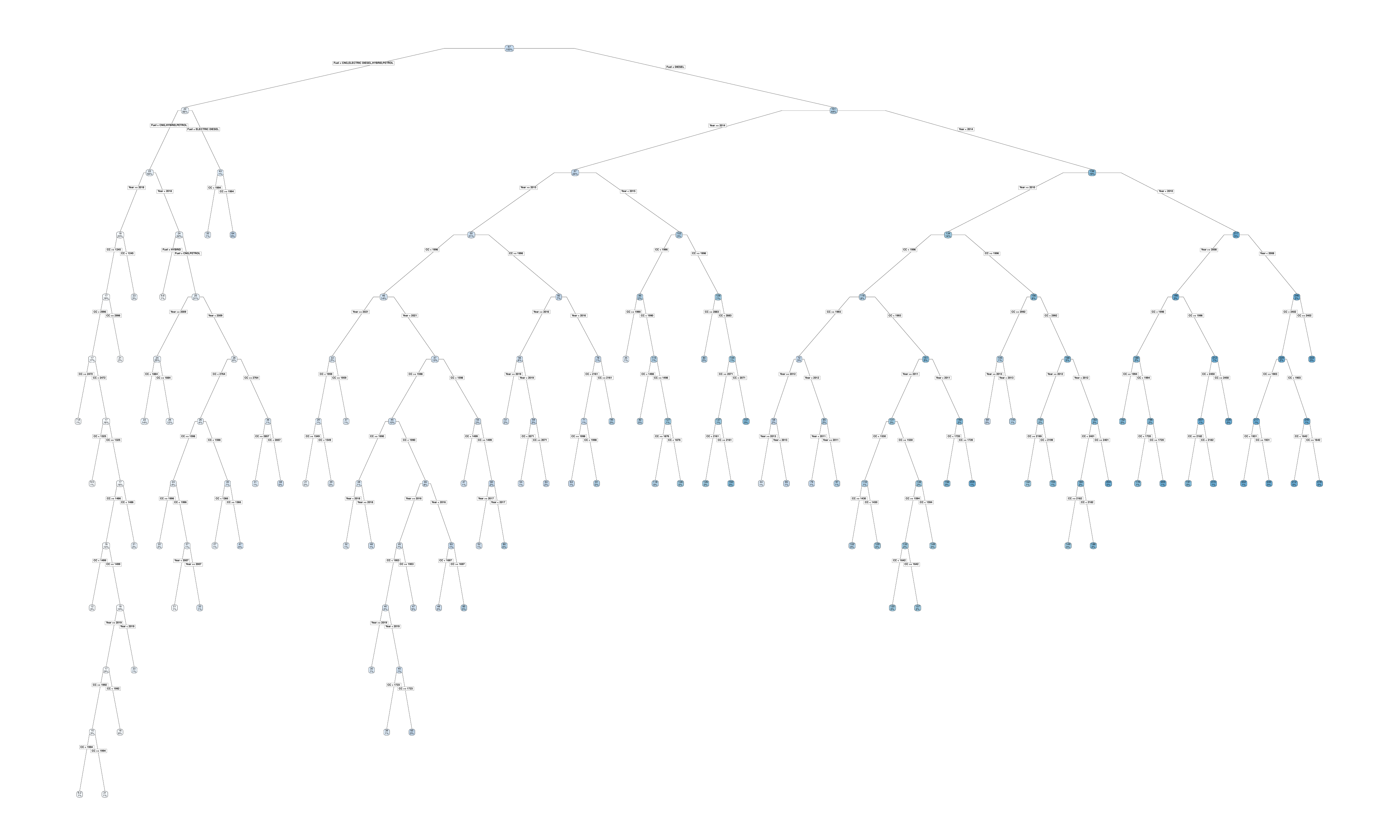}
    \caption{Recursively Partitioned Decision Tree: NOx}
    \end{center}
\end{figure}
\end{landscape}
\end{document}